\documentclass[12pt]{iopart}

\usepackage{iopams}

\usepackage{graphicx}

\begin{document}

\title{Co-existence of ferromagnetic and antiferromagnetic interactions in Mn$_3$Ga$ _{(1-x)} $Sn$_x$C}

\author{E.T. Dias$^1$, K.R. Priolkar$^1$, A.K. Nigam$^2$}

\address{$^1$ Department of Physics, Goa University, Goa 403206, India}
\address{$^2$ Tata Institute of Fundamental Research, Dr. Homi Bhabha Road, Colaba, Mumbai 400005, India}
\ead{krp@unigoa.ac.in}

\begin{abstract}The magnetic properties of the Mn$_3$Ga$_{(1-x)}$Sn$_x$C, 0 $\le$ x $\le$ 1 antiperovskite compounds have been investigated in detail. Though all compounds of this series crystallize in a cubic structure, the Ga rich (x $\le$ 0.2) compounds transform, via a first order transformation, to an antiferromagnetic ground state and the Sn rich (x $\ge$ 0.8) compounds exhibit dominant ferromagnetic interactions at low temperature. In the intermediate range (0.4 $\le$ x $\le$ 0.7) co-existence of ferromagnetic and antiferromagnetic interactions can be seen. The results have been explained to be due to growth of ferromagnetic sub-lattice at the expense of antiferromagnetic sub-lattice with increasing Sn concentration in Mn$_3$Ga$_{(1-x)}$Sn$_x$C. This growth occurs to a point where the first-order transition is altered from a ferromagnetic - antiferromagnetic type in Ga rich compounds to a paramagnetic - ferrimagnetic type in Mn$_3$SnC.

\end{abstract}

\maketitle

\section{Introduction}
Since the report of superconductivity in MgCNi$ _{3} $ \cite{Gs1}, antiperovskite compounds of type M$ _{3} $AX (M = Mn, Fe, Ni; A = Ga, Al, Sn, Zn, Cu, In, Ge, Ag; X = C, N) have been extensively investigated \cite{Gs2,Gs3,Gs4,Gs5,Gs6,Gs7,Gs8,Gs9,Gs10,Gs11,Gs12,Gs13}. Interest in Mn based materials of this type in recent times has been due to properties like giant magnetoresistance (GMR) \cite{Gs14}, negative thermal expansion (NTE) \cite{Gs15,Gs16} and near zero temperature coefficients of the resistivity (NZ-TCR) \cite{Gs12,Gs20}. The wide range in physical properties suggests possibilities of a wider range of chemical bonding via partial replacement of constituent elements that could control the magnetostructural correlations in these materials \cite{Gsa}.

Among these, Mn$  _{3}$GaC has attracted considerable attention due to its multiple magnetic transitions. With decreasing temperature it exhibits three phase transitions: first, a second-order transition from paramagnetic (PM) to a collinear ferromagnetic (FM) phase at Curie temperature (T$_{C}$) $\sim249K$. The next transition is to a canted ferromagnetic (CFM) phase at T $\sim 164K$ via a second-order transition \cite{Gs28} and the third is a first-order transition to the antiferromaagnetic (AFM) ground state at $\sim160K$ \cite{Gs28,Gs29} accompanied by a discontinuous change in volume \cite{Gs30} and a 5K thermal-hysteresis \cite{Gsr}. Contrastingly, the compound Mn$_{3}$SnC presents a spontaneous magnetization change from the room temperature PM state to a ferrimagnetic (FIM) one via a first-order transition around T$ _{C}$ \cite{Gss}.
Results of neutron diffraction suggest that the compound possesses a complicated spin arrangement consisting of AFM and FM sub-lattices perpendicular to each other resulting in a non-collinear FIM state \cite{Gs2,Gst,Gsyy}. Unlike its Ga containing counterpart, Mn$_3$SnC undergoes only one magnetic transition from PM to FIM state which also involves an abrupt change in its unit cell volume \cite{Gss,Gst,Gsu,Gsv}.

Various attempts to understand properties accompanying the first-order transition have been made by altering the external conditions such as pressure and dopants \cite{Gsv,Gsae}. Effect of pressure on Mn$  _{3}$GaC is to stabilize the FM phase as observed through increase of T$_C$ and a concomitant decrease in first-order transformation temperature \cite{Gsae}. Further, high pressure X-ray diffraction studies up to 35 GPa revealed no structural transition \cite{Gse}.

Transition metal substitutions at the Mn sites result in weakening the first-order transformation. Fe doping in Mn$  _{3}$SnC not only gives rise to a competition between the AFM and FM components in the FIM state resulting in a broadened FM region but also quenches the NTE behavior reported in Mn$ _{3} $SnC \cite{Gsn}.
Ni or Co doping in Mn$_3$GaC results in a gradual decrease in the AFM first-order transition temperature \cite{Gsb,Gsc}. In Ni-doped compounds, the MR exceeds 60\%-70\% at 5T which is attributed to the weakened AFM ground state.

Variation of carbon content has the most dramatic effects. Increasing the concentration of carbon in Mn$_{3}$SnC$ _{x}$ leads to a growth of FM sub-lattice at the expense of AFM sub-lattice \cite{Gsq} while the carbon deficiency in the compound Mn$_{3}$GaC$ _{x} $ results in a total disappearance of the first-order AFM ground state at low temperatures \cite{Gsz,Gszz}.

Strong correlations between the metalloid (C,N) and its octahedron of surrounding Mn atoms allows the substitution of various metals at the A-site \cite{Gs27}. Ge-doping in Mn$ _{3} $X$ _{(1-y)} $Ge$ _{y} $C (X = Al, Zn, Ga, Sn) causes local lattice distortions which in turn provokes a local magnetic disorder reducing the first-order magnetic transition temperature. Ge doping in Mn$ _{3} $SnC gradually reduces the MVE, till it disappears. Substitution of Al for Ga in Mn$ _{3} $GaC produces the same effect as the application of high pressure while partial substitution of Zn increases T$  _{C}$ and maximal values of MR along with gradually suppressing the AFM ground state.

From the above it is clear that doping suppresses or changes the character of first-order magnetic transformation indicating the importance of correlations between Mn and other constituents. Mn$_3$GaC and Mn$_3$SnC are two compounds which exhibit a contrasting evolution of their properties. While Mn$_3$GaC has an AFM ground state, its Sn counterpart orders ferrimagnetically. With different doping FM state is seen to stabilize in Mn$_3$GaC, AFM interactions strengthen in Mn$_3$SnC. There is no clear understanding of transformation from an antiferromagnetic ground state to a ferrimagnetic one as Ga is replaced by Sn. A critical balance of ferromagnetic and antiferromagnetic interactions is required for the materials applicability in magnetic refrigeration. While ferromagnetic ordering is responsible for large changes in magnetic entropy, antiferromagnetic order suppresses hysteresis. Therefore it would be interesting to study the evolution of the first-order magnetic transformation in the solid solutions of the type Mn$_3$Ga$_{(1-x)}$Sn$_x$C, $0 \le x \le 1$.

\section{Experimental}
In order to synthesize polycrystalline samples of general formula Mn$ _{3} $Ga$ _{(1-x)} $Sn$ _{x} $C, (x=0, 0.1, 0.2, 0.4, 0.5, 0.6, 0.8, 0.9 \& 1.0), desired proportions of the starting materials were weighed, mixed and pressed into pellets before annealing them in evacuated quartz tubes at 1073K for the first 48 hours and 1150K for 120 hours \cite{Gsyy}. About 15 wt.\% excess carbon was added to realize in stoichiometric carbon compounds \cite{Gszz,Gsxx}. After cooling to room temperature the phase formation and purity of the samples were examined by recording X-ray diffraction (XRD) patterns using Cu-K$\alpha$ radiation. Studies of the magnetic properties of the samples were carried out in a Quantum Design Magnetic Property Measurement System (MPMS) in the temperature range 5K to 390K and in applied magnetic fields of $0.01T$ and $0.5T$ and between $ \pm 7T $. Magnetization as a function of temperature was recorded under three experimental modes: zero-field cooling (ZFC), field-cooled-cooling (FCC), and field-cooled warming (FCW) in an applied magnetic field of 0.01T. For the ZFC mode, each sample was cooled down to 5K from room temperature in the remanant field ($\sim$ 0.001T) before applying the necessary magnetic field. The ZFC data was then recorded while warming. The subsequent cooling and warming magnetization curves with applied field on, are termed as FCC and FCW respectively. Isothermal magnetization at different temperatures in the field range of $\pm 7T$ were recorded in ZFC mode. Similarly electrical resistivity of the samples was measured by standard four probe method using a Quantum Design Physical Properties Measurement System (PPMS) in the temperature range 5K and 300K during cooling and warming cycles.

\section{Results}
\begin{figure}[h]
\centerline{\includegraphics[width=\textwidth]{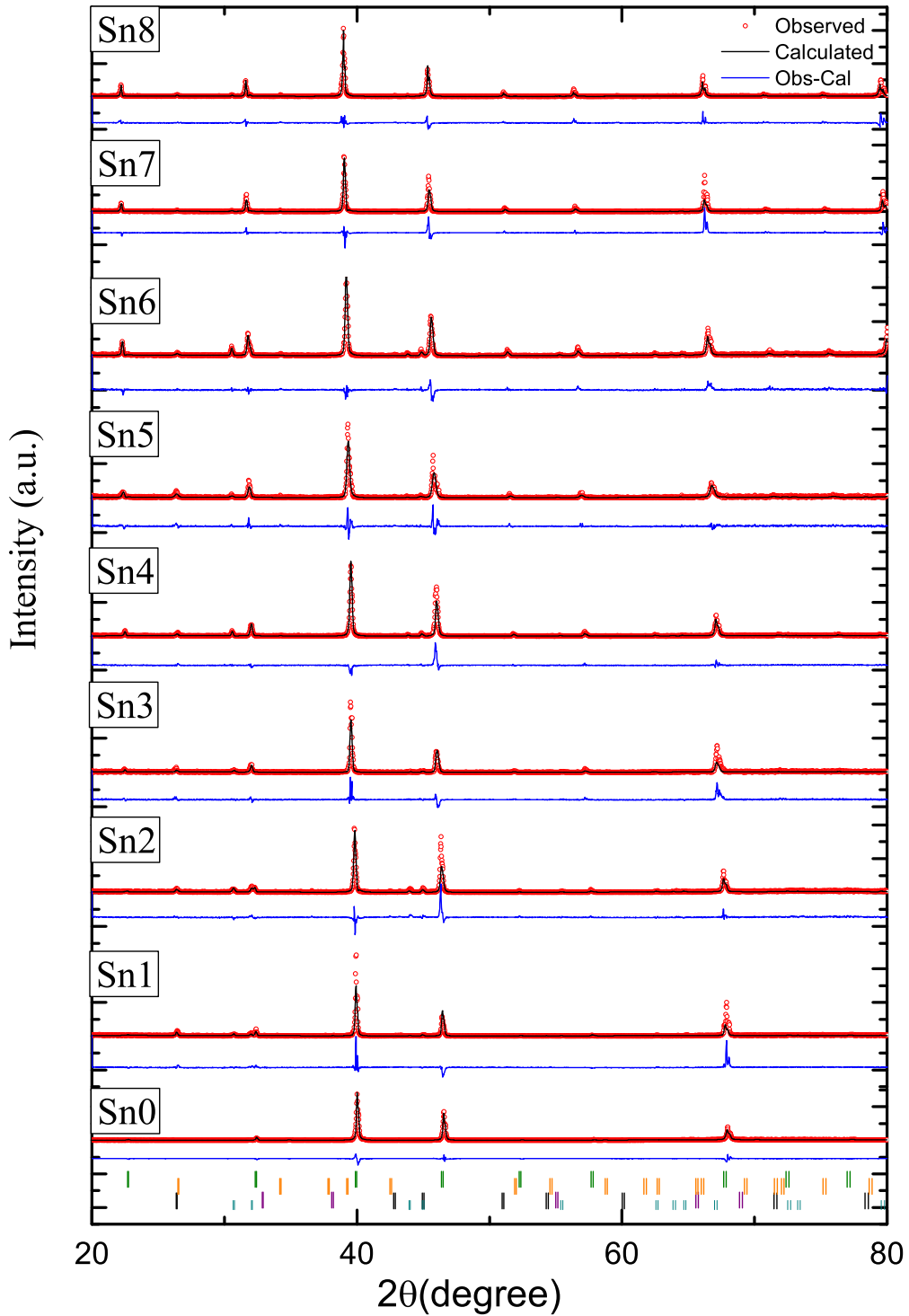}}
\caption{Room temperature X-ray diffraction patterns of Mn$ _{3} $Ga$ _{(1-x)} $Sn$ _{x} $C, $0 \le x \le 1$ refined using Rietveld method.} \label{fig:xrd}
\end{figure}

X-ray diffraction patterns recorded in the $ 20^\circ \leq 2\theta \leq 80^\circ$ range were Rietveld refined using FULLPROF suite. Refinement revealed that the XRD patterns could be well indexed using the structural model (space group: \textit{Pm$\bar{3}$m}) reported for Mn$ _{3} $GaC as shown in Figure \ref{fig:xrd} along with minor impurity phases of MnO ($\sim 1\%$) and SnO ($ < 1\%$) Sn($\sim 2\%$) and graphite (maximum $\sim 5\%$). Along with global parameters like, scale factor and background, lattice parameter and Sn/Ga ratio were refined for the major phase. The refined concentrations of Sn at Ga site are listed in Table \ref{tbl:riet1}. It can be seen that compounds Sn1 and Sn2 have similar values of Sn concentration and so is the case with Sn6 and Sn7 which could be due to volatility of Ga.

\begin{table}
\caption{Values of Sn concentration obtained from Rietveld refinement of X-ray diffraction patterns.}
\label{tbl:riet1}
\begin{center}
\begin{tabular}{|c|c|c|}
\hline\hline
Sample Name & Nominal composition & Refined value \\
\hline
Sn0 & $x$=0 & $x$=0 \\
\hline
Sn1 & $x$=0.1 & $x$=0.23 $\pm0.02$ \\
\hline
Sn2 & $x$=0.2 & $x$=0.20 $\pm0.02$ \\
\hline
Sn3 & $x$=0.4 & $x$=0.41$ \pm0.03 $ \\
\hline
Sn4 & $x$=0.5 & $x$=0.55$ \pm0.01 $ \\
\hline
Sn5 & $x$=0.6 & $x$=0.71$ \pm0.02 $ \\
\hline
Sn6 & $x$=0.8 & $x$=0.97$ \pm0.02 $ \\
\hline
Sn7 & $x$=0.9 & $x$=0.92$ \pm0.02 $ \\
\hline
Sn8 & $x$=1.0 & $x$=1.0 \\
\hline\hline
\end{tabular}
\end{center}
\end{table}

\begin{figure}
\centerline{\includegraphics[width=\textwidth]{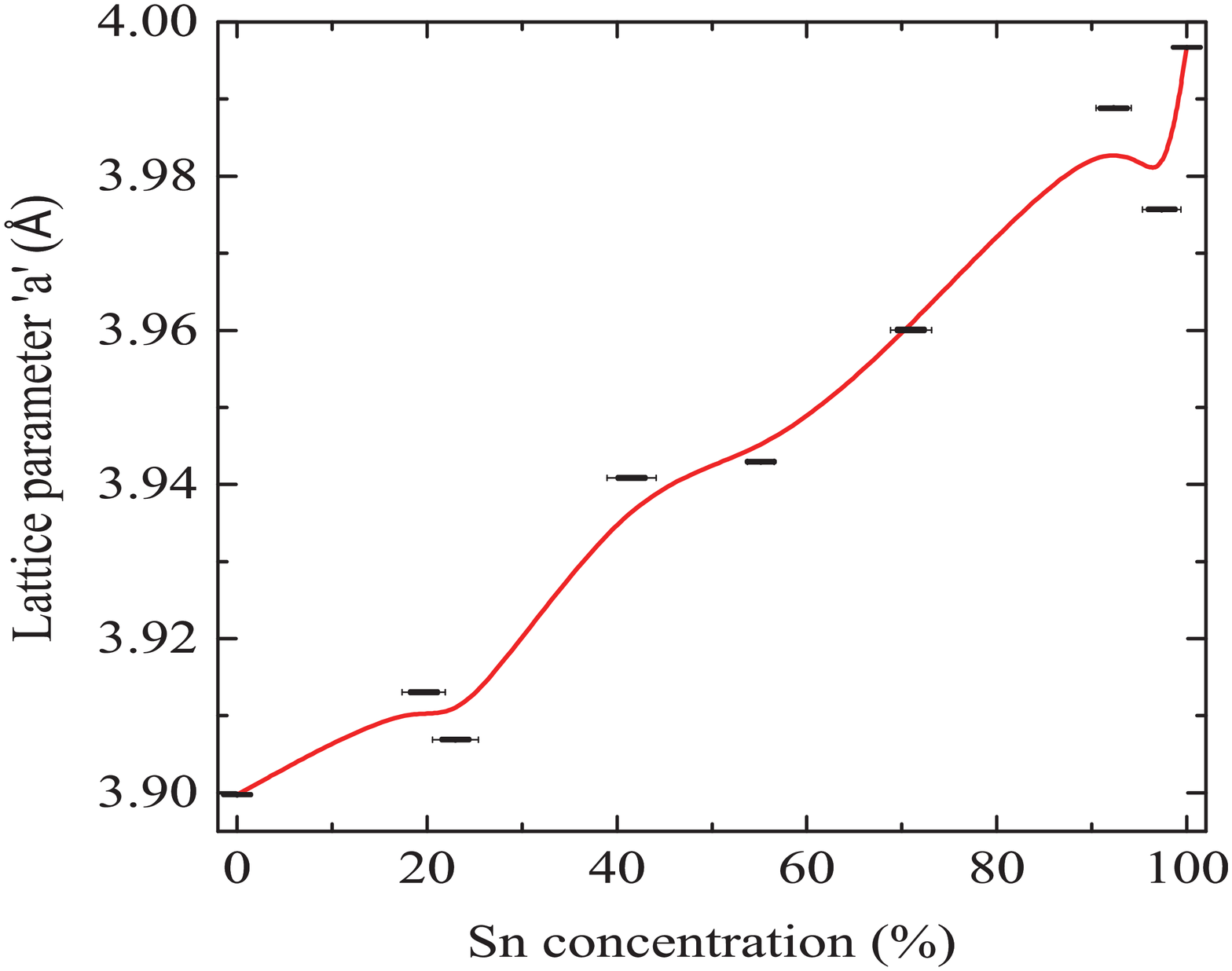}}
\caption{Variation of lattice constant $a$ as a function Sn concentration in Mn$ _{3} $Ga$ _{(1-x)} $Sn$ _{x} $C, $0 \le x \le 1$.} \label{fig:lattice}
\end{figure}

A plot of lattice constants as a function of Sn concentration in Figure \ref{fig:lattice} with values obtained from refinement shows a systematic increase across the entire composition range. This result is consistent with the fact that Sn atoms are larger than those of Ga and successfully replace them at the A-site. Such a replacement by Sn results in an increased cell volume with increase in doping concentration of Sn \cite{Gsn,Gsab} thus obeying Vegard's law.

\begin{figure}[htb]
\centerline{\includegraphics[width=\textwidth]{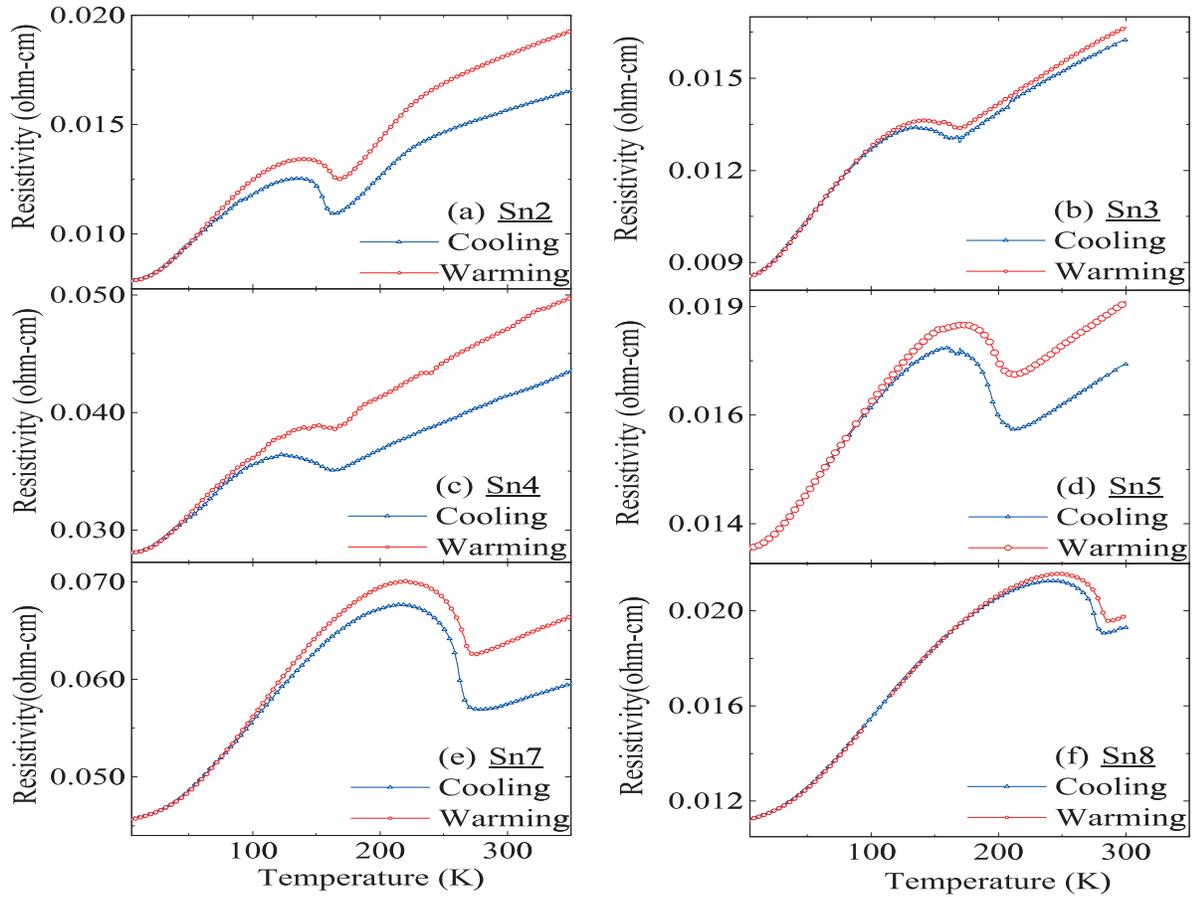}}
\caption{Plots of resistivity measured at zero field as a function of temperature for Mn$_3$Ga$_{1-x}$Sn$_x$C, $0 \le x \le 1$ }
\label{fig:res1}
\end{figure}

Electrical resistivity measurements for the Mn-based antiperovskites show a metallic behaviour over the entire temperature range. Figure \ref{fig:res1} compares temperature dependent resistivity plots for the Mn$ _{3} $Ga$ _{(1-x)} $Sn$ _{x} $C series of compounds measured during the cooling and warming cycles. Temperature dependence of resistivity for all the samples is quite similar.  Figure \ref{fig:res1}a shows the variation of resistivity in Sn2 compound. A distinct change in slope is observed at about $220K$ followed by a minimum and a sharp rise thereafter at T $\sim 160K$. In case of Mn$_3$GaC, these temperatures indicate second and first-order magnetic transitions respectively \cite{Gs28,Gs29}. The strong discontinuity and hysteresis observed between cooling and warming curves all along the series suggests that the compounds in the series undergo a first-order type of transition. With increase in Sn concentration although overall metallic behavior remains the same, the minimum indicating first-order transition peak exhibits a gradual shift towards higher temperatures. Another change observed in resistivity curves is decrease in sharpness of the resistivity increase after the minimum in case of Sn3 and Sn4 compounds. The sharpness of this transition returns for compounds with higher Sn concentration. In the case of the other end member the series, Sn8 (Mn$_3$SnC), the anomaly in resistivity at about T $\sim 279K$, coincides with  PM to FIM transformation \cite{Gsyy}.

In order to understand the nature of both first-order and second-order transformations in Mn$ _{3} $Ga$ _{(1-x)} $Sn$ _{x} $ magnetization measurements were carried out. The results of these measurements allow us to classify the compounds in the following three categories:

\begin{enumerate}
\item{Gallium rich compounds (Sn0, Sn1 \& Sn2)}
\item{Intermediate compounds (Sn3, Sn4 \& Sn5) each of which will be discussed separately}
\item{Tin rich compounds (Sn6, Sn7 \& Sn8)}
\end{enumerate}

\subsection{Gallium rich compounds}
\begin{figure}
\centerline{\includegraphics[width=\textwidth]{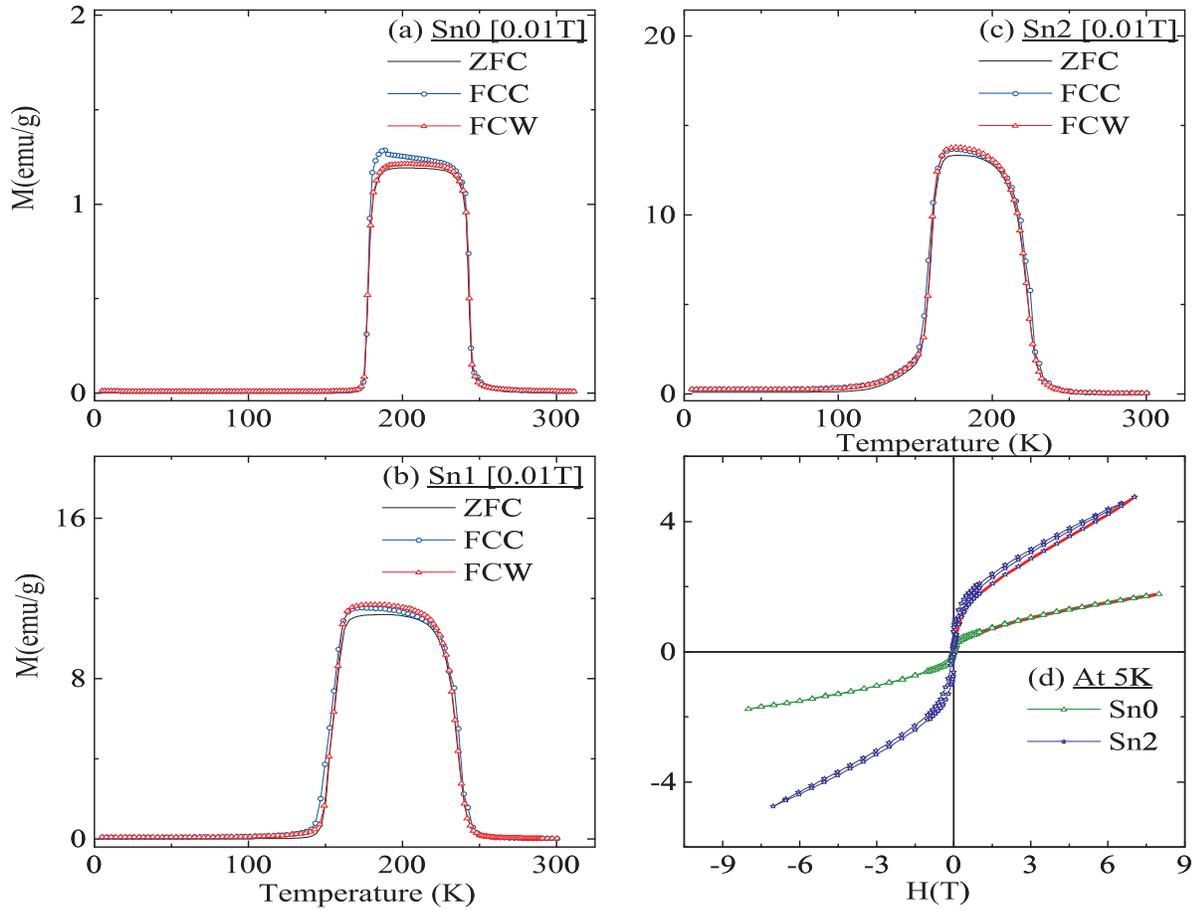}}
\caption{Magnetization vs. Temperature M(T) curves for (a). Sn0 (b). Sn1 (c). Sn2 during ZFC, FCC and FCW cycles at H=0.01T. (d). Isothermal magnetisation curves M(H) recorded at 5K for Sn0 and Sn2 (Initial magnetization curves highlighted in a different colour).}
\label{fig:mtga}
\end{figure}

Figure \ref{fig:mtga} depicts magnetization curves recorded in ZFC, FCC and FCW modes in applied field of $0.01T$. Magnetization as a function of temperature M(T) of Mn$ _{3}$GaC (Sn0) exhibits a second-order transition from the high temperature PM phase to a FM one at T$ _{C} $ = $243K$ followed by a first-order transformation to AFM state at T$_N$ $\sim178K $. As expected, the M(T) plots for the samples Sn1 and Sn2 are similar to those of Sn0, exhibiting a PM - FM transition at T$ _{C}$ $\sim236K $ and T$ _{C}$ $\sim222K $ respectively. The first-order transformations to AFM ground state are also seen at T$_N$ $\sim155K$ and T$ _{N} $ $\sim161K $ respectively for Sn1 and Sn2 samples. These are lower than the first-order transformation temperature of Sn0 indicating Sn doping in Mn$_3$GaC atleast in Ga rich compounds decreases both the first-order and second-order transformation temperatures.

\begin{figure}
\centerline{\includegraphics[width=\textwidth]{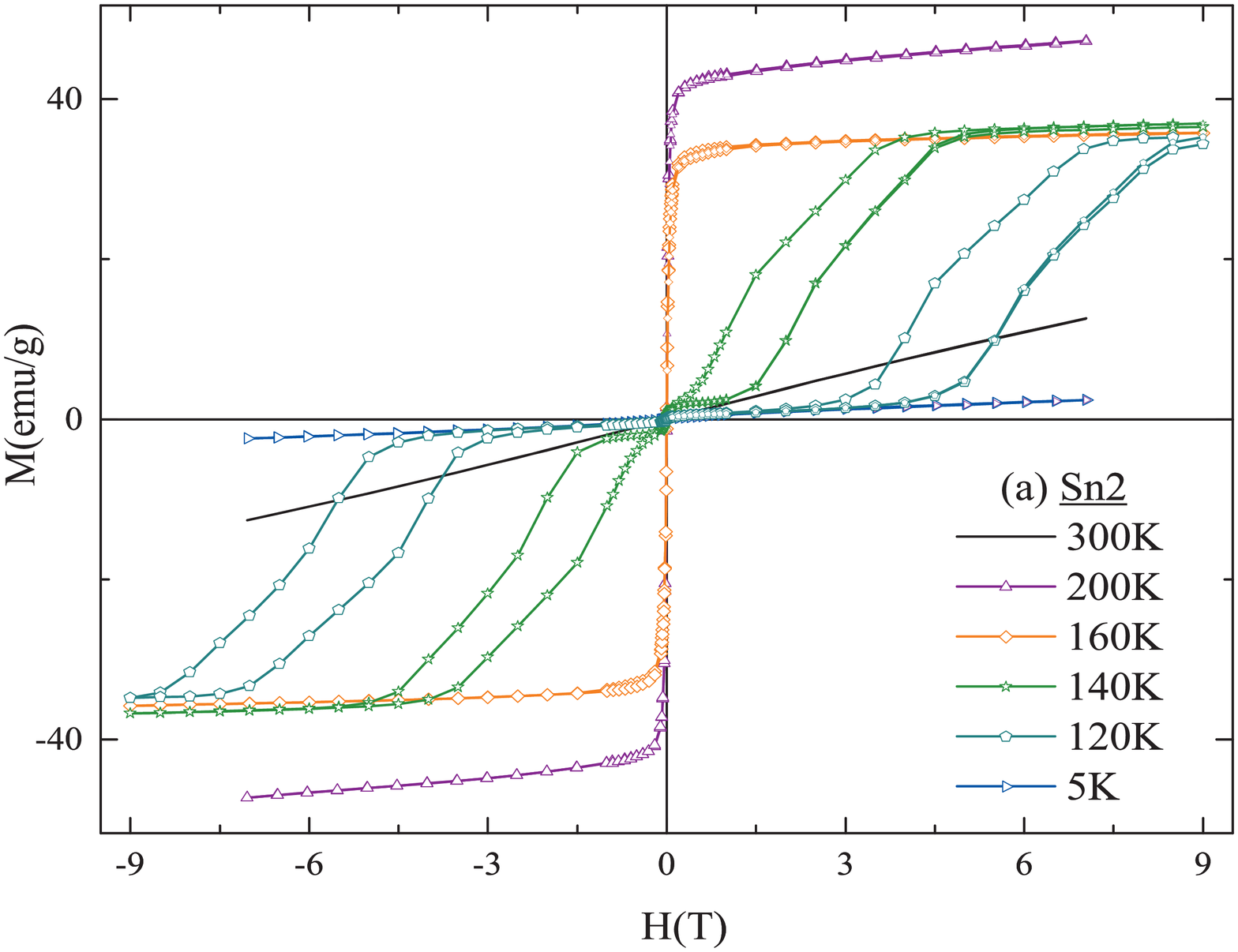}}
\caption{M(H) plots recorded at selected temperatures for compound Sn2.}
\label{fig:mhsn1}
\end{figure}

In order to understand the effect of Sn in greater detail, isothermal magnetization curves were recorded and studied at selected temperatures in the $ \pm 7T$ field range. A comparison of hysteresis loops recorded at $5K$ for Sn0 and  Sn2 are shown in Figure \ref{fig:mtga}d. Although both the compounds are AFM at $5K$, M(H) loops exhibit a curvature but with non saturating magnetic moment even in a field of $7T$. The magnetization increases noticeably with Sn replacing Ga in the Sn2 compound. Another distinct feature of the hysteresis loop is the initial magnetization curve lying outside the loop. This  suggest the existence FM interactions in the compound even at $5K$. To further understand the magnetic interactions in Sn doped compound, the isothermal magnetic response at select temperatures were recorded and are plotted in Figure \ref{fig:mhsn1}. While the $300K$ loop depicts paramagnetic behaviour, the loops at $200K$ and $160K$ indicate ferromagnetic ordering of the sample. AFM ground state of the sample is clear from the hysteresis loop recorded at $5K$ with interfering FM interactions as shown earlier. The M(H) loops recorded just below the AFM transformation, show a clear metamagnetic transition from AFM to FM state. It can also be seen that the closer one is to the transformation temperature, the AFM - FM transition occurs in lower applied fields. Such a field induced transition has been also seen in Mn$_3$GaC near its transformation temperature indicating that the magnetic interactions in these low Sn concentration samples are similar to those in the undoped Mn$_3$GaC. Effect of Sn doping at these lower concentrations seems to introduce ferromagnetic interactions which can be clearly seen in M(H) loop recorded at T = $5K$. It also lowers the first-order FM - AFM transformation temperature.

\subsection{Intermediate compounds}
\begin{figure}
\centerline{\includegraphics[width=\textwidth]{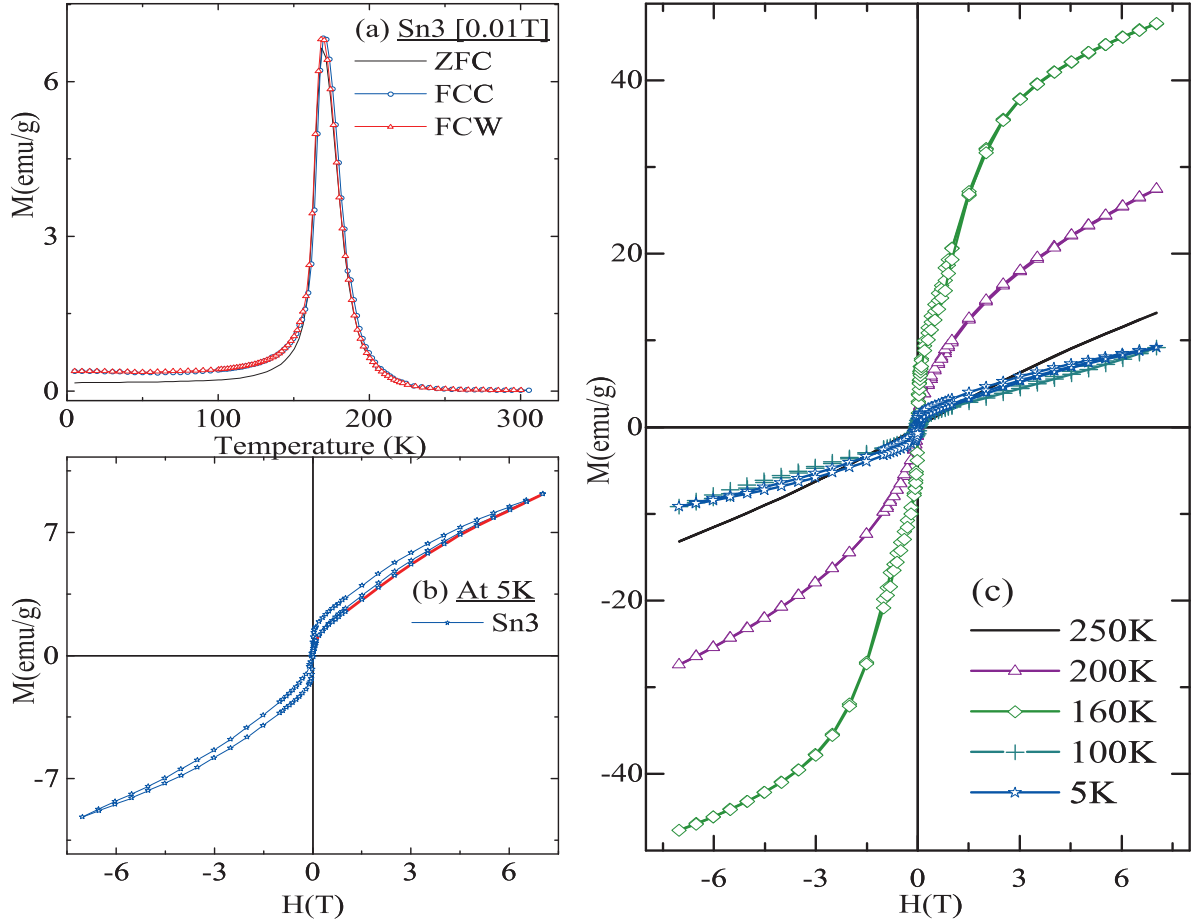}}
\caption{(a). M(T) curves recorded for Sn3 after ZFC and during FCC and FCW cycles at H=0.01T. (b). Isothermal magnetisation curve recorded for Sn3 at 5K (Initial magnetization curve highlighted in a different colour). (c). Comparison of M(H) plots recorded for Sn3 at selected temperatures.} \label{fig:sn3}
\end{figure}

Here we discuss magnetization measurements on three intermediate compositions Sn3, Sn4 and Sn5 individually. Temperature dependent magnetization of Sn3 presented in Figure \ref{fig:sn3}a reveals an interesting picture. Contrary to expectations that increasing Sn doping would support ferromagnetic order at the expense of antiferromagnetism, the ferromagnetic ordering temperature of Sn3 compound decreases while the antiferromagnetic transformation temperature increases as compared to Sn2 compound. Furthermore, it appears that even before the  FM transition is complete the compound abruptly transforms to the AFM state via a first-order transition at T $\sim 164K$. However, the hysteresis loops in Figure \ref{fig:sn3}b. and c. recorded at specific temperatures depict that although the compound has dominant AFM interactions, FM correlations are also present. The hysteresis loops recorded at all temperature $ \le$ $200K$ have a non saturating $\lq$S' shaped curvature indicating presence of competing FM and AFM interactions. The hysteresis loop recorded at T = $160K$, which is just below its transformation  temperature exhibits a weak metamagnetic transition.  At $5K$ the virgin magnetization curve lies outside the envelope hinting at presence of strongly competing FM interactions. From the hysteresis loops, it appears that with decreasing temperature, both FM and AFM interactions strengthen.

\begin{figure}
\centerline{\includegraphics[width=\textwidth]{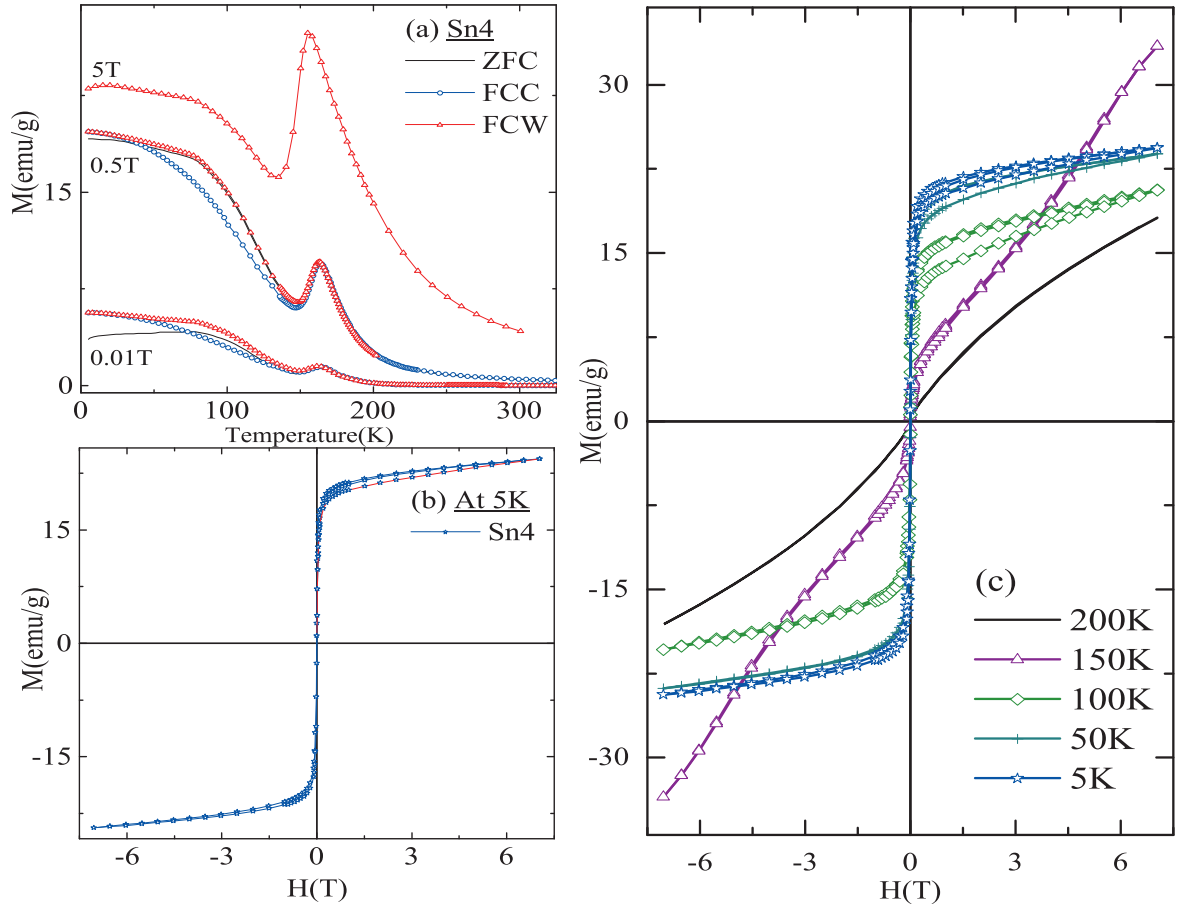}}
\caption{{\bf a.} Variation of M(T) curves for Sn4 after ZFC and during FCC and FCW cycles at H=0.01T, 0.5T and 5T. (b). Plot of M(H) at 5K for Sn4 (Initial magnetization curves highlighted in a different colour). (c). Temperature variations of M(H) curves for Sn4 compound.}
\label{fig:sn4}
\end{figure}

At about 50\% doping of Sn for Ga, the magnetization results are even more interesting. The M(T) curves recorded in ZFC, FCC and FCW modes for Sn4 compound are displayed in Figure \ref{fig:sn4}. Magnetization recorded in a applied magnetic field of $0.01T$ exhibits a relatively sharp upturn indicating some kind of ferromagnetic ordering at about T $ \sim 170K$. Absence of hysteresis between the FCC and FCW magnetization curves indicates the transition to be of second-order. However, as in Sn3 even before the transformation to ferromagnetic state is complete, magnetization drops sharply only to rise again even more strongly below $155K$. Both these transitions appear to be first-order in nature as indicated by presence of hysteresis between FCC and FCW magnetization curves. It appears as if one part of the sample tends to favour antiferromagnetic alignment of the Mn spins during structural deformation, while the other part facilitates FM order. One could also argue that the compound is essentially ferromagnetic along with magnetic inhomogeneities. If this be the case, then higher magnetic field should favor ferromagnetic alignment and smear out the signatures of antiferromagnetic ordering seen at around $160K$. However, both FM and AFM transitions are present in magnetization recorded at H = $0.5T$ and $5T$. Furthermore, both these transformation temperatures reduce with increasing magnetic field.  Another confirmation for presence of FM and AFM regions is obtained from isothermal magnetization recorded at select temperatures. Presence of a hysteresis loop at $5K$ indicates the compound to be ferromagnetic as also indicated by M(T) curves. However, the non saturation magnetic moment up to H = $7T$ is indicative  of presence of antiferromagnetic interactions. At higher temperatures ($50K$, $100K$), the M(H) loops indicate weakening ferromagnetism and a gradual built up of antiferromagnetic interactions. At $150K$, the M(H) loop is of a typical antiferromagnet but with a metamagnetic transition. Such metamagnetic transitions are seen in all Ga rich samples and are typical of Mn$_3$GaC. These observations lend weight to the argument that with increasing Ga concentration, AFM regions grow at the expense of FM regions.

\begin{figure}
\centerline{\includegraphics[width=\textwidth]{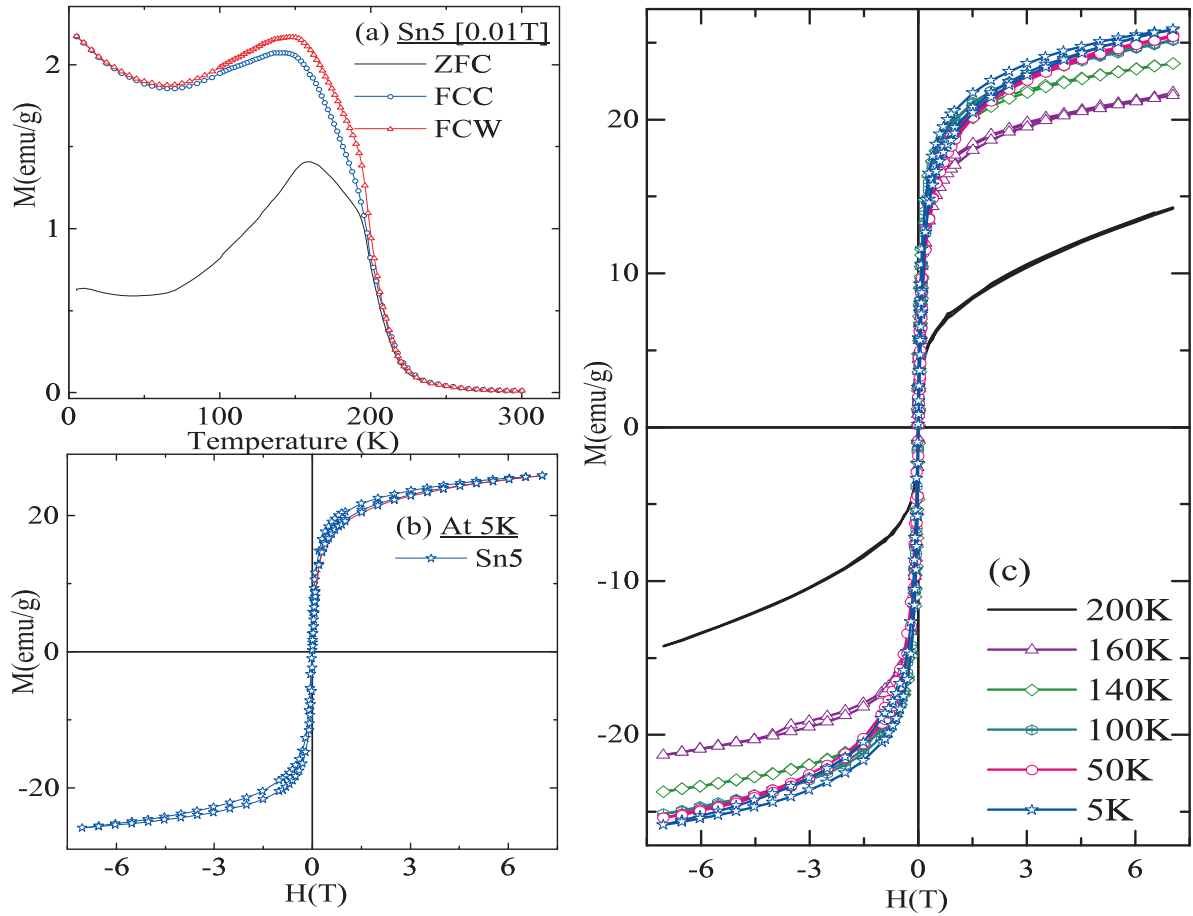}}
\caption{{\bf a.} Plots of M(T) for Sn5 recorded after ZFC and FCC and FCW cycles at H=0.01T. (b). Field dependent magnetization curves for Sn5 compound at 5K and (c). at and below T$ _{C} $} \label{fig:sn5}
\end{figure}

Results of M(T) measurements carried out on Sn5 compound (Figure \ref{fig:sn5}) show that as the concentration of tin is further increased to about $ \sim$70\%, the compound transforms from a PM state to a FM like ordering via a first-order transition at T$ _{C} $ $ \sim 198K$. However, there are other characteristics that also dominate the magnetization behavior. Firstly, there is a large divergence between ZFC and FC curves indicating presence of FM and AFM regions. Secondly, the magnetization peaks at about $150K$ and then passes through a minimum at about $75K$. Presence of hysteresis between FCC and FCW magnetization curves in this entire temperature range indicates that the alignment of Mn spins is continuously evolving through out the first-order transformation. It is known that Mn$_3$SnC consist of FM and AFM sublattices at low temperatures \cite{Gs2,Gst,Gsyy}. Decrease of carbon content in Mn$_3$SnC increases AFM regions at the expense of FM regions. In Sn5 compound, it appears that the presence of nearly 30\% Ga forces Mn spins to also align antiferromagnetically. This is further accentuated by M(H) curves. Hysteresis loop recorded at $5K$ indicates dominance of FM interaction however, the magnetic moment does not saturate even in magnetic fields up to $7T$. The M(H) curves recorded in the range of $50K$ $\le$ T $\le$ $200K$ exhibit similar characteristics. The value of magnetization increases with decreasing temperature except below $100K$ where a reversal of trend is noticed. While the magnetization values recorded at $50K$ and $100K$ are similar, those at  $5K$ are distinctly smaller. This confirms the presence of strongly competing ferro and antiferromagnetic regions.

\subsection{Tin rich compounds}
\begin{figure}
\centerline{\includegraphics[width=\textwidth]{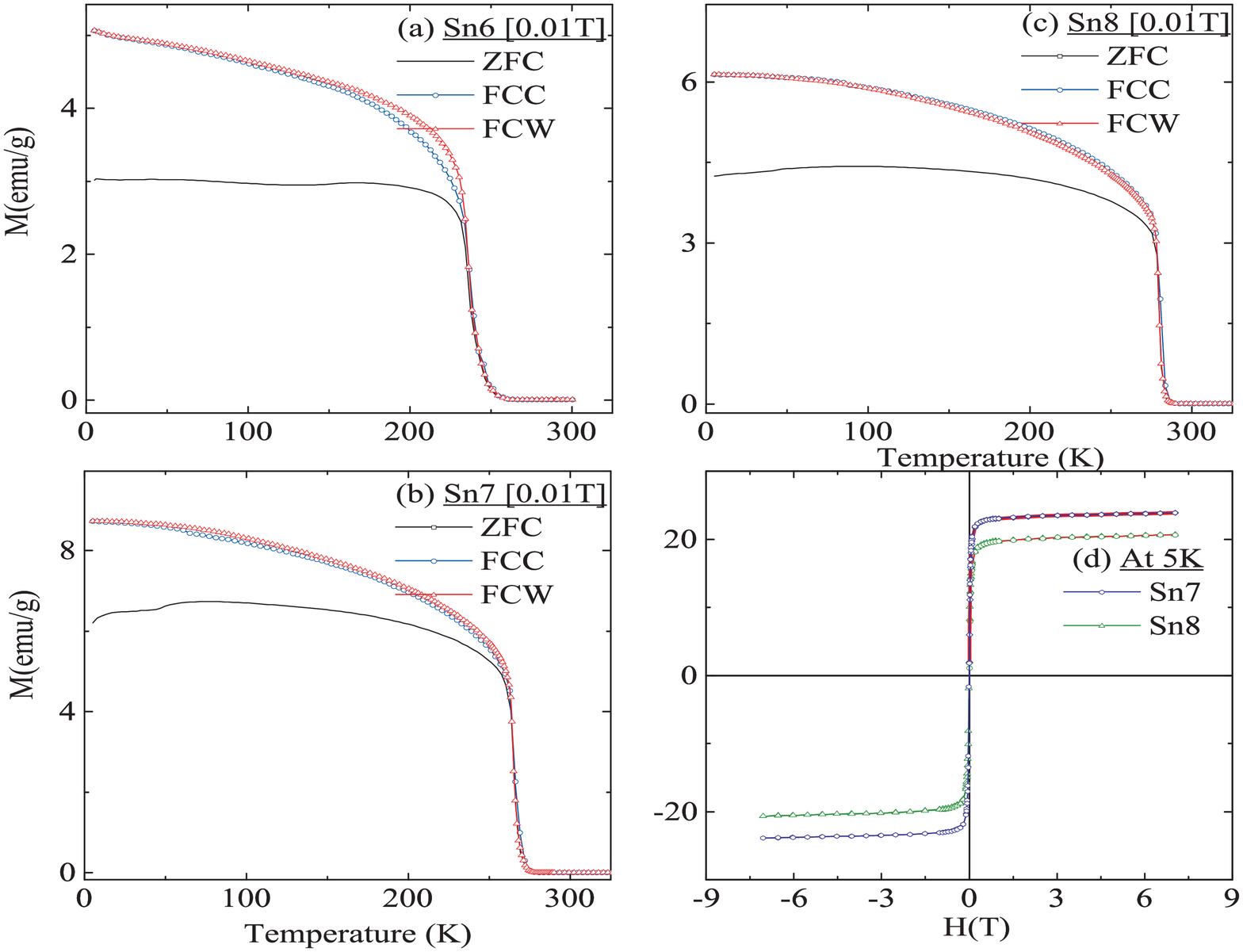}}
\caption{a. M(T) plots for Sn6, Sn7 and Sn8 recorded after ZFC and during FCC and FCW cycles at H=0.01T. d. M(H) plots for Sn7 and Sn8 recorded at 5K (Initial magnetization curves highlighted in a different colour).}
\label{fig:mtsn}
\end{figure}

\begin{figure}
\centerline{\includegraphics[width=\textwidth]{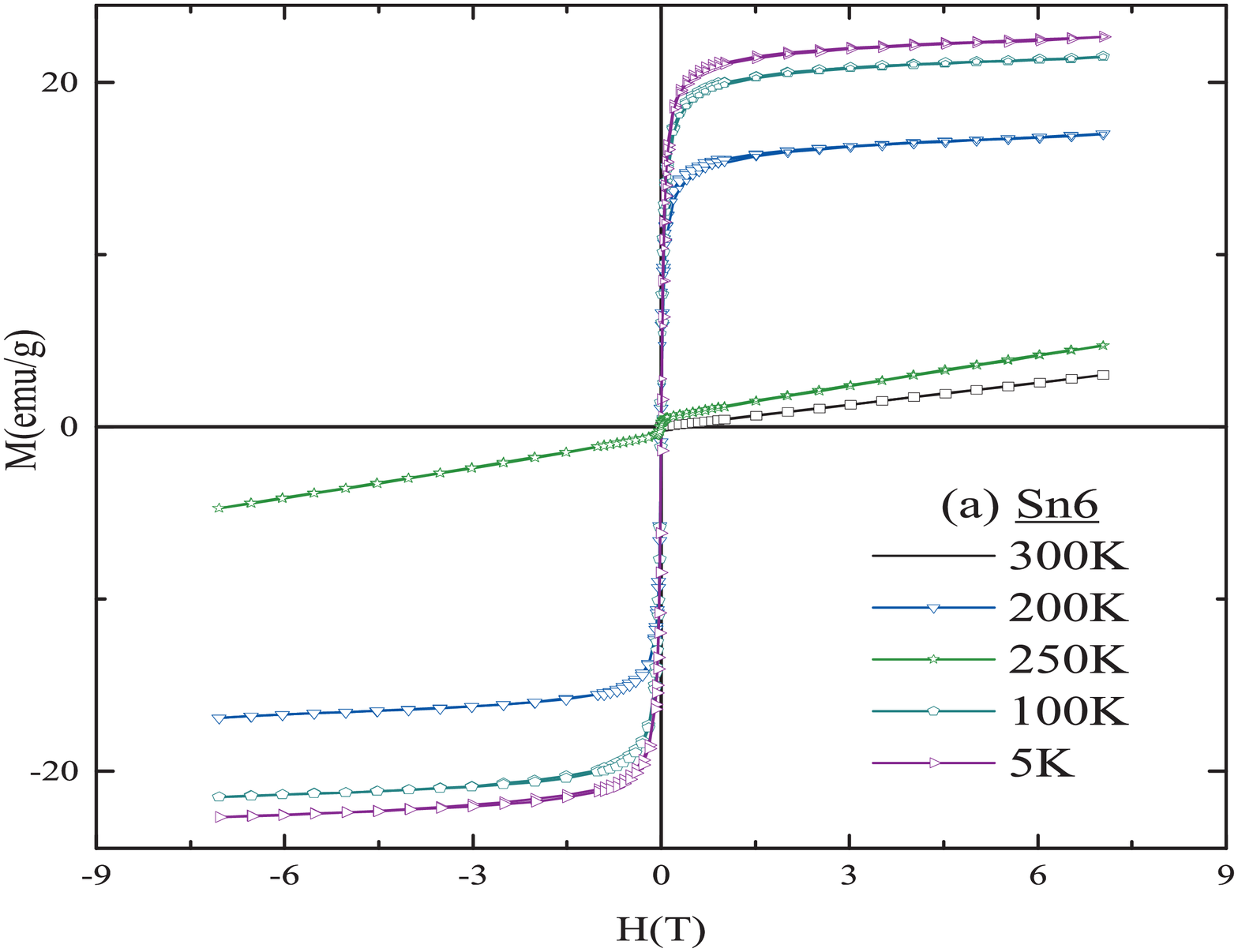}}
\caption{M(H) plots for Sn6 at various temperatures above and below the first-order transition.}
\label{fig:mhsn6}
\end{figure}

Results of magnetization measurements carried out on tin rich compounds Sn6, Sn7 \& Sn8 are shown in Figure \ref{fig:mtsn} and Figure \ref{fig:mhsn6}.
Mn$_3$SnC (Sn8) has a ferrimagnetic ground state and exhibits a single first-order transformation from paramagnetic state at about $\sim 279K$. With increasing Ga doping, this transformation temperature decreases to about $250K$. Another noticeable feature is the increasing difference in magnetization recorded in ZFC and FC modes with increasing Ga doping. This difference tends to diminish in magnetization curves recorded in higher magnetic fields and hence it can be attributed to magnetic inhomogeneities introduced by Ga doping.

Magnetic hysteresis loops recorded at $5K$ for Sn7 and Sn8 are presented in Figure \ref{fig:mtsn}d. The two loops are expectedly very similar to each other as the high fields wash out the effect of magnetic inhomogeneities. For both samples, magnetization increases steeply and tends to saturate above $0.5T$ demonstrating a typical soft ferromagnetic character. Field-dependent magnetization curves at select temperatures above and below T$ _{C} $ = $236K$ for Sn6 compound are plotted in Figure \ref{fig:mhsn6}. M(H) loop recorded at $300K$ and $250K$ are in paramagnetic state. While those recorded below $T_C$ confirm the presence of ferromagnetic ordering.

\section{Discussion}

\begin{figure}
\centerline{\includegraphics[width=\textwidth]{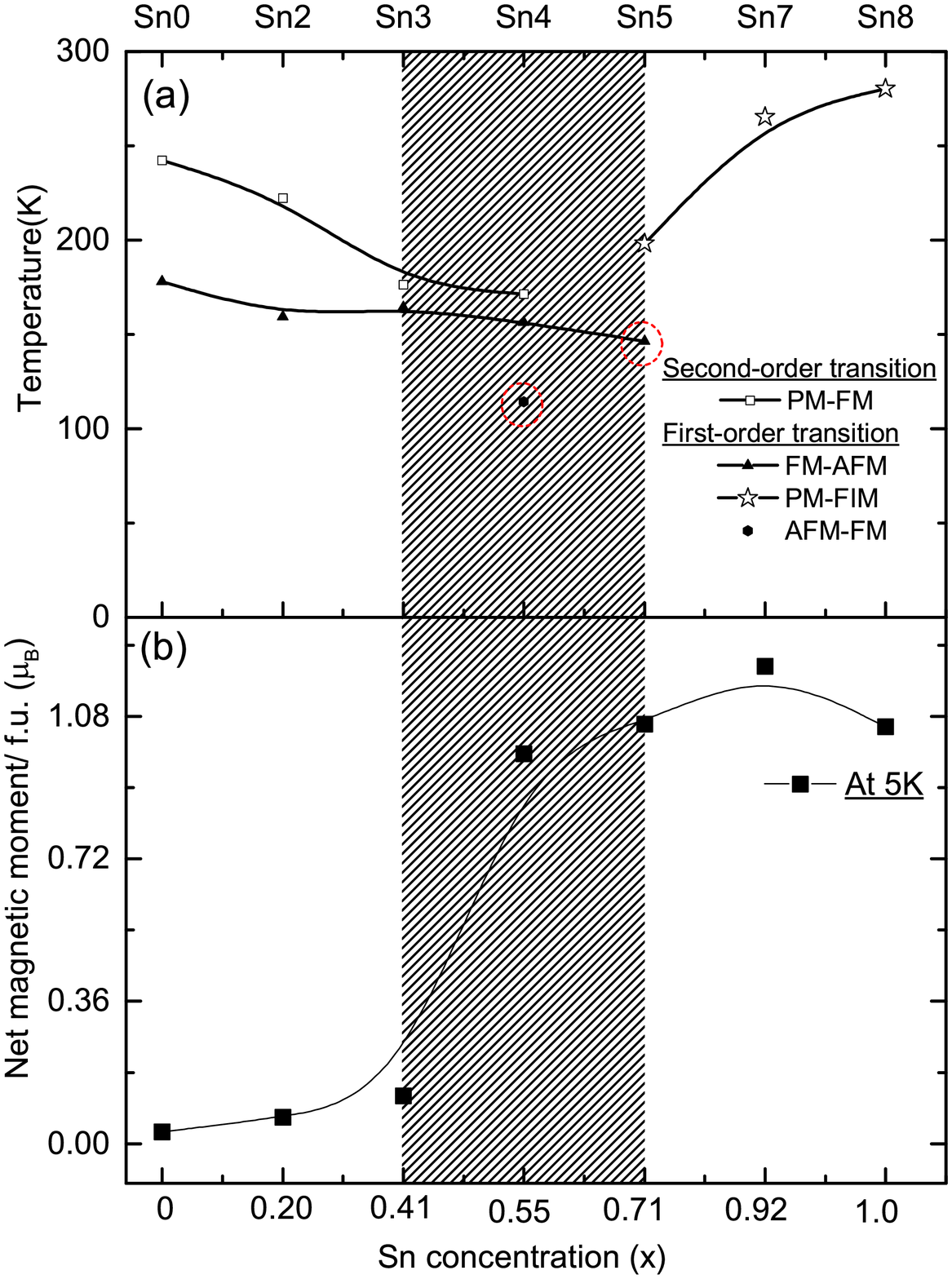}}
\caption{a. Transition temperatures as a function Sn concentration for the system Mn$ _{3} $Ga$ _{(1-x)} $Sn$ _{x} $C, $0 \le x \le 1$ system (AFM= antiferromagnetic phase, FM=ferromagnetic phase FIM=Ferrimagnetic phase and PM=paramagnetic phase). b. Variation of net magnetic moment per formula unit with increasing concentrations of Sn.}
\label{fig:phase}
\end{figure}

In Mn$_3$Ga$_{(1-x)}$Sn$_x$C compounds, while the Ga rich compounds have an antiferromagnetic ground state, the Sn rich compounds exhibit dominance of ferromagnetic interactions over antiferromagnetic ones. In the intermediate range, with increasing Ga concentration, a growth of AFM regions at the expense of FM regions can be seen from magnetization measurements. This is also supported by the variation of various transition temperatures and the magnetic moment values obtained from M(H) curves at $5K$. The transition temperatures of the first-and second-order magnetic transformations for the Mn$ _{3} $Ga$ _{(1-x)} $Sn$ _{x} $C compounds were determined from the first derivative of the ZFC curves. As illustrated in Figure \ref{fig:phase}a, all the Ga rich samples exhibit a second order paramagnetic to ferromagnetic transition before undergoing a magneto-structural transformation to an antiferromagnetic state. The $T_C$ of this second order transition decreases with increasing Sn concentration before disappearing or merging into the first-order transformation curve. The first-order transformation for Ga rich samples is from a ferromagnetic state to an antiferromagnetic state. Again with decreasing Ga content, this transformation temperature too decreases, though very gradually and again rises quite steeply in the Sn rich region ($x \ge 0.5$). However, for these compounds, the transformation is from a paramagnetic state to a ferrimagnetic state as reported in case of Mn$_3$SnC.  This indicates that the Ga rich samples, ($x < 0.5$) basically have antiferromagnetic ground state but with presence of ferromagnetic interactions. These ferromagnetic interactions could be field induced as in case of Mn$_3$GaC or could be due to presence of Sn. The Sn rich samples on the other hand ($x \ge 0.5$) have ferrimangetic ground state which is a mixture of ferromagnetic and antiferromagnetic interactions with antiferromagnetic interactions dominating with increasing Ga doping. This is again clear from a plot of net magnetic moment per formula unit estimated from M(H) curves recorded at $5K$. This net magnetic moment in each compound was obtained by linear extrapolation of the magnetization value at H = 7T to meet the magnetization axis at H = 0T. These values plotted in Figure \ref{fig:phase}b. It can be seen that Ga rich compounds have much lower magnetic moment values as compared to Sn rich compounds. The magnetic moment values also show a sudden jump around $x = 0.5$, which seems to indicate a change in magnetic order. However, these intermediate range compounds ($0.4 \le x \le 0.7$) exhibit multiple magnetic transitions and are discussed separately.

In Mn$_3$Ga$_{0.45}$Sn$_{0.55}$C, it can be seen that there are two magnetostructural transformation. It seems as if there exist two magnetic phases which are ordering independently of each other. It is plausible that the regions that is rich in Ga order antiferromagnetically while those rich in Sn order ferrimagnetically. The existence of two magnetic phases can be also seen in compounds with  $x$ = 0.4 and $x$ = 0.7. In the first case ($x$ = 0.4), the Ga rich phase first tends to order ferromagnetically via a second order transition and then orders antiferromagnetically through a magnetostructural transformation. This transformation temperature is lower than that in Mn$_3$GaC or 20\% Sn doped sample. However, even in the antiferromagnetic phase, there are ample signatures of existence of ferromagnetic correlations. A comparison of  M(H) loops recorded at 5K in Sn0, Sn2 and Sn4 shows a systematic increase in magnetization and coercive field with increasing Sn concentration. These are signatures of stronger ferromagnetism. The situation is completely reversed in Mn$_3$Ga$_{0.3}$Sn$_{0.7}$C. Here the compound first orders ferromagnetically via a first-order transformation as in case of Mn$_3$SnC or 20\% Ga doped compound but at lower temperatures, the magnetization suddenly dips giving indication of an antiferromagnetic transformation. The presence of hysteresis between magnetization values recorded while warming and cooling indicates the transformation to be of first-order in nature. Again the M(H) loops recorded at lower temperatures show a decrease in values of magnetization below 50K. The initial magnetization curves recorded at 50K and 5K are seen to be lying outside the hysteresis loop. These are again clear signatures of presence of two competing magnetic phases.  

Therefore the magnetization studies carried out on Mn$_3$Ga$_{1-x}$Sn$_x$C type antiperovskites show a evolution of first-order magnetic transformation from ferromagnetic - antiferromagnetic type in Ga rich compounds to paramagnetic - ferrimagnetic type in Sn rich compounds. In the intermediate doping range, the magnetization studies indicate co-existence of two magnetic phases: Ga rich phase which orders antiferromagnetically and a Sn rich phase with dominant ferromagnetic interactions. 

\section{Conclusions}
In conclusion, magnetization of Mn$_3$Ga$_{1-x}$Sn$_x$C, $0 \le x \le 1$ series of compounds indicates that though Sn doping does not alter the cubic structure or metallic behavior of the compounds it gradually increases the strength of ferromagnetic interactions across the series while suppressing the antiferromagnetic ground state. The Ga-rich compounds ($x \le 0.2$) transform to an antiferromagnetic ground state via a first-order transition but with a decreasing first order transition temperature with Sn doping. A similar behavior is noticed in case of Sn rich compounds ($x \ge 0.8$) wherein the para to ferrimagnetic ordering temperature decreases as Ga is doped in Mn$_3$SnC. In the intermediate concentration range ($0.4 \le x \le 0.7$), a co-existence of ferromagnetic and antiferromagnetic order is seen in their magnetic properties. The results indicate that as Sn is doped at the Ga-site in Mn$_3$Ga$_{1-x}$Sn$_x$C, ferromagnetic sub-lattice grows at the expense of antiferromagnetic sub-lattice to a point where the first-order transition is altered from a FM-AFM type in Ga rich compounds to a PM-FIM type in Mn$_3$SnC.

\section{Acknowledgments}
Authors thank Board of Research in Nuclear Sciences (BRNS) for the financial support under the project 2011/37P/06. M/s Devendra D. Buddhikot and Ganesh Jangam are acknowledged for the experimental assistance.

\end{document}